\documentclass[conference,10pt]{IEEEtran}
\IEEEoverridecommandlockouts

\usepackage{url}
\usepackage{cite}
\usepackage{tikz}
\usepackage{braket}
\usepackage{lipsum}
\usepackage{subfig}
\usepackage{xcolor}
\usepackage{amsmath}
\usepackage{balance}
\usepackage{environ}
\usepackage{physics}
\usepackage{colortbl}
\usepackage{enumitem}
\usepackage{amsfonts}
\usepackage{booktabs}
\usepackage{graphicx}
\usepackage{textcomp}
\usepackage{tcolorbox}
\usepackage{algorithm}
\usepackage{algpseudocode}
\usepackage{arydshln}
\usepackage[normalem]{ulem}
\usepackage{siunitx}
\usepackage{makecell}

\tcbuselibrary{skins}
\tcbuselibrary{breakable}

\newcounter{RQcounter}
\newcommand\showRQcounter{\stepcounter{RQcounter}\theRQcounter}

\newcounter{TQcounter}
\newcommand\showTQcounter{\stepcounter{TQcounter}\theTQcounter}

\newtcolorbox{insightbox}[3][]
{
  breakable,
  enhanced,
  colback=orange!50!gray!10, 
  colframe=red!80!black, 
  boxrule=0.6pt,
  left=2mm,
  right=2mm,
  top=2mm,
  bottom=1.5mm,
  title={},
  overlay={
    \node[
      fill=red!80!black,     
      text=white, 
      font=\bfseries\itshape, 
      anchor=south west, 
      rounded corners=2pt, 
      inner xsep=8pt,         
      inner ysep=4pt          
    ] 
    at ([xshift=10pt,yshift=-7pt]frame.north west)
    {Observation \showTQcounter\ #2};
  },
  before skip=6pt,
  after skip=6pt,
  #1,
}

\def\BibTeX{{\rm B\kern-.05em{\sc i\kern-.025em b}\kern-.08em
    T\kern-.1667em\lower.7ex\hbox{E}\kern-.125emX}}
\begin{document}

\title{Revisiting Noise-adaptive Transpilation in Quantum Computing: How Much Impact Does it Have?\vspace{5mm}}

\author{\IEEEauthorblockN{Yuqian Huo}
\IEEEauthorblockA{\textit{Rice University}, USA}
\and
\IEEEauthorblockN{Jinbiao Wei}
\IEEEauthorblockA{\textit{Rice University}, USA}
\and
\IEEEauthorblockN{ Christopher Kverne}
\IEEEauthorblockA{\textit{Florida International University (FIU)}, USA}
\and
\IEEEauthorblockN{Mayur Akewar}
\IEEEauthorblockA{\textit{Florida International University (FIU)}, USA}
\and
\IEEEauthorblockN{Janki Bhimani}
\IEEEauthorblockA{\textit{Florida International University (FIU)}, USA}
\and
\IEEEauthorblockN{Tirthak Patel}
\IEEEauthorblockA{\textit{Rice University}, USA}
}

\maketitle

\begin{abstract}

Transpilation, particularly noise-aware optimization, is widely regarded as essential for maximizing the performance of quantum circuits on superconducting quantum computers. The common wisdom is that each circuit should be transpiled using up-to-date noise calibration data to optimize fidelity. In this work, we revisit the necessity of frequent noise-adaptive transpilation, conducting an in-depth empirical study across five IBM 127-qubit quantum computers and 16 diverse quantum algorithms. Our findings reveal novel and interesting insights: (1) noise-aware transpilation leads to a heavy concentration of workloads on a small subset of qubits, which increases output error variability; (2) using random mapping can mitigate this effect while maintaining comparable average fidelity; and (3) circuits compiled once with calibration data can be reliably reused across multiple calibration cycles and time periods without significant loss in fidelity. These results suggest that the classical overhead associated with daily, per-circuit noise-aware transpilation may not be justified. We propose lightweight alternatives that reduce this overhead without sacrificing fidelity -- offering a path to more efficient and scalable quantum workflows.


\end{abstract}


\pagestyle{plain}

\section{Introduction}
\label{sec:introduction}

Quantum computing has the potential to revolutionize fields like cryptography, optimization, and machine learning by leveraging quantum phenomena such as superposition, entanglement, and interference~\cite{castelvecchi2024ai,preskill2022quantum}. However, near-term quantum hardware, such as superconducting qubits, remains highly susceptible to noise, resulting in errors that degrade computational accuracy~\cite{smith2022scaling,kjaergaard2020superconducting}. To mitigate these issues, quantum computing frameworks like IBM's Qiskit provide \textit{transpilation}~\cite{aleksandrowiczqiskit}, a process that optimizes circuits by mapping qubits and gates to specific hardware configurations, taking into account noise and connectivity constraints. 

Over the last few years, noise-aware transpilation has become a widely adopted strategy to improve quantum circuit fidelity on noisy quantum hardware~\cite{endo2021hybrid,tannu2019ensemble,tannu2019not,murali2020software,zulehner2019compiling,liu2023tackling,tan2023compiling,patel2020ureqa,patel2020veritas}. Transpilation is also predicted to remain an important aspect of quantum circuit compilation, even in the early fault-tolerant quantum computing (FTQC) era, because of the qubit heterogeneity of quantum systems, which will require careful transpilation to find computer sections with low enough error rates to operate in the error-correction regime\cite{murali2019noise,li2019tackling,tannu2019not}. In addition, recent studies have analyzed noise effects and transpilation efficacy across various quantum computing systems~\cite{ravi2021quantum,liu2020reliability,patel2020experimental,ash2020experimental,dahlhauser2021modeling}, noting the reliance on constantly updated calibration data. These studies underscore the perceived importance of frequent re-transpilation to counter temporal drift in hardware noise characteristics.

However, these works were primarily conducted three to four years ago, using algorithms requiring fewer qubits ($<10$) and less advanced quantum devices. Larger circuits face more complex error propagation patterns. Modern 127-qubit devices present different connectivity and noise profiles. At the same time, the classical overhead of transpilation -- especially under full noise-aware optimization -- has increased considerably, often adding seconds per circuit, which scales poorly for workloads requiring thousands of circuit runs, such as variational algorithms~\cite{endo2021hybrid}. Yet it remains unclear if such frequent and costly recompilation is still necessary on today’s hardware.

\textit{Our study aims to investigate if the assumed importance of transpilation still holds for the largest superconducting quantum computers for a diverse set of complex algorithms\footnote{This work is published in the Proceedings of the International Conference on Computer-Aided Design (ICCAD), 2025.}.} We empirically evaluate the necessity and impact of frequent noise-aware transpilation using a month-long study of 16 quantum algorithms (size of up to 31 qubits) on six 127-qubit IBM quantum computers, investigating whether the classical overhead associated with continual re-transpilation is justified in the context of the modern hardware landscape. Conducting such a study is non-trivial: quantum computers offer limited daily access, calibration data changes frequently, and empirical fidelity evaluation requires repeated executions over time. Nevertheless, our month-long open-source effort is designed to probe whether the conventional wisdom behind frequent noise-aware transpilation continues to hold in practice.

\vspace{2mm}

\noindent\textbf{Our results reveal some novel and insightful findings:}

\vspace{2mm}
    
\begin{enumerate}[leftmargin=*]
    \item Noise-adaptive transpilation often concentrates usage on a small subset of ``good'' qubits, resulting in uneven resource utilization and increased output error variability due to dynamic qubit wear and fluctuations.

    \vspace{1mm}
    
    \item While circuit routing techniques improve the output fidelity, noise-adaptive mapping techniques are often not impactful -- in fact, we find that simple random mapping achieves comparable average fidelity with significantly less output fluctuation, and also avoids qubit overuse.

    \vspace{1mm}

    \item Using lower transpiler optimization levels (e.g., levels 1 or 2) incurs far less compilation time and yields similar output fidelity compared to the most aggressive optimization level (level 3), especially for larger circuits.

    \vspace{1mm}

    \item We observe no statistically significant correlation between the time of execution and output fidelity, indicating that qubit drift over the course of a calibration cycle has minimal practical impact on algorithm results.

    \vspace{1mm}
    
    \item Circuits compiled once with calibration data from a single day maintain their fidelity when executed across multiple calibration cycles, challenging the conventional wisdom that frequent re-transpilation is necessary.
\end{enumerate}

\vspace{2mm}

Based on these insights, we propose simple and lightweight methods to reduce transpilation overhead and mitigate output error, advancing more efficient approaches to executing quantum algorithms on superconducting devices. Our methods and data are open-sourced at: \\\textit{\url{https://github.com/positivetechnologylab/ICCAD25}}.
\section{Background and Motivation}
\label{sec:background}

\begin{figure}[t]
    \centering
    \includegraphics[width=0.98\columnwidth]{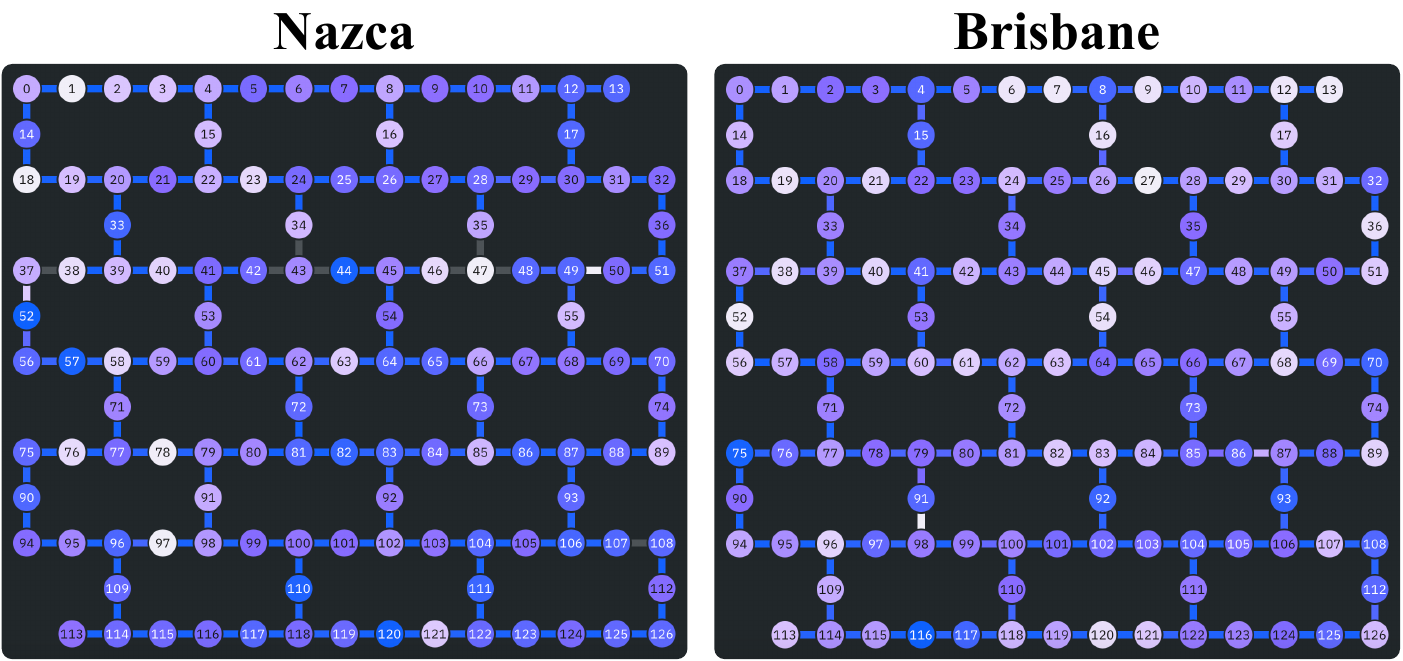}
    \vspace{1.5mm}
    \hrule
    \vspace{1.5mm}
    \caption{Example 127-qubit IBM Nazca and Brisbane quantum computers. The circles represent the qubits (color-coded according to their respective T1 times), and the lines represent the qubit connections (color-coded as per ECR gate errors). Darker color indicates better values.}
    \vspace{-3mm}
    \label{fig:computers}
\end{figure}

Quantum computing harnesses the principles of quantum mechanics to process information in ways fundamentally different from classical computers. Unlike classical bits, which can only be in one of two states (0 or 1), quantum bits, or qubits, can exist in a superposition of both 0 and 1 simultaneously. This property, combined with quantum entanglement and interference, enables quantum computers to perform certain calculations (gates and operations) far more efficiently than their classical counterparts.

\begin{figure*}[t]
    \centering
    \includegraphics[width=0.99\textwidth]{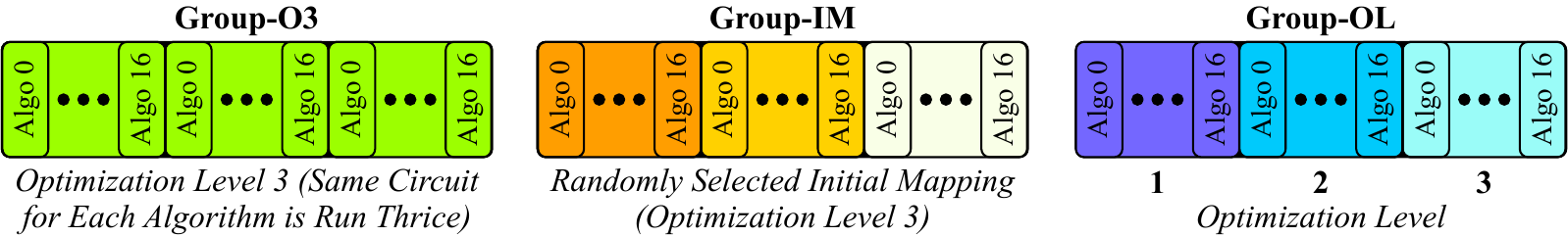}
    \vspace{1.5mm}
    \hrule
    \vspace{1.5mm}
    \caption{We run three groups of experiments (all three groups are run daily), where each group is assembled into one job with $16\times{}3=48$ circuits. Here, 16 is the number of algorithms run, and we run each cluster of algorithms thrice within each group. All groups are submitted in parallel on each available computer daily. If a group's job does not finish within the calibration cycle for which it is optimized, the job is canceled or disregarded. Groups may end up running in any order. Each color in the visualization represents a specific transpiled circuit variation according to the mapping methodology or optimization level.}
    \vspace{-3mm}
    \label{fig:grouping}
\end{figure*}

\subsection{Superconducting Qubits and Noise Effects}

Superconducting qubits, the most advanced qubit realization currently, are fabricated using materials that exhibit zero electrical resistance when cooled to extremely low temperatures, enabling stable quantum states~\cite{kjaergaard2020superconducting,wilen2021correlated}. These qubits are manipulated using microwave pulses to implement quantum gates. However, due to the inherent fragility of quantum states, superconducting qubits are highly susceptible to noise, which can introduce errors into quantum computations~\cite{ravi2021quantum,dahlhauser2021modeling}. Noise manifests in several forms, each characterized by distinct error metrics as described below.

\vspace{1mm}

\begin{itemize} [leftmargin=*]
    \item \textbf{Relaxation Time ($T_1$):} The $T_1$ time is the time it takes for an excited qubit to decay to its ground state.

    \vspace{2mm}

    \item \textbf{Dephasing Time ($T_2$):} The $T_2$ time measures the duration over which a qubit maintains its phase coherence.

    \vspace{2mm}

    \item \textbf{One- and Two-Qubit Gate Errors:} Gate operations are prone to control inaccuracies, which lead to gate errors. These errors accumulate as gate operations are applied within a circuit, impacting overall fidelity.

    \vspace{2mm}

    \item \textbf{Readout Errors:} Quantum computations are finalized by measuring qubit states, but this measurement process can introduce errors. Readout errors occur when a qubit in state 0 or 1 is incorrectly measured as the opposite state.
\end{itemize}

\vspace{2mm}

The noise characteristics of superconducting qubits vary not only across different qubits (shown for IBM quantum computers in Fig.~\ref{fig:computers}) but also over time due to changes in the physical environment and device qubit drift. IBM quantum computers, for example, are calibrated daily to minimize these noise effects. Consequently, the noise parameters ($T_1$, $T_2$, gate error, readout error, and CX error) are subject to temporal variations, which complicates the task of achieving reliable quantum computations on such hardware~\cite{patel2020experimental,liu2020reliability}.

\subsection{Transpilation and Optimization Levels in Qiskit}

Quantum circuits designed for an ideal, noise-free model must be adapted to the constraints of real quantum hardware through a process known as \textit{transpilation}. Transpilation maps an abstract quantum circuit onto the physical qubits and gates available on a quantum processor, taking into account the connectivity and noise characteristics of the device. The transpiler in Qiskit~\cite{aleksandrowiczqiskit}, IBM's widely-used quantum computing framework, provides three levels of optimizer passes implemented from prior, state-of-the-art research~\cite{li2019tackling,tucci2005introduction}:

\vspace{1mm}

\begin{itemize}[leftmargin=*]
    \item \textbf{Level 1: Minimal Optimization} \textit{(Basic Routing and Error Reduction)} - At this level, the transpiler applies only basic routing transformations to ensure that all required qubit connections are physically feasible (e.g. use swap gates to move qubits adjacent to each other when needed). Level 1 prioritizes speed.

    \vspace{2mm}

    \item \textbf{Level 2: Moderate Optimization} \textit{(Balanced Error Mitigation)} - In addition to basic routing, Level 2 introduces more aggressive error-mitigation techniques. This level performs further gate cancellation and circuit rewriting, balancing the trade-off between transpilation time and circuit fidelity.

    \vspace{2mm}

    \item \textbf{Level 3: Full Optimization} \textit{(Advanced Noise-Aware Optimization)} - This level provides the most thorough optimization, considering noise characteristics and targeting specific low-error qubits for placing critical circuit paths whenever possible. Level 3 is the most time-intensive.
\end{itemize}

\subsection{Motivation for our Study}

While noise-aware transpilation is widely accepted as a best practice in today’s quantum computing workflows, this assumption has not been thoroughly reevaluated in the context of modern, large-scale quantum hardware. Much of the foundational work in this space was conducted on earlier systems with tens of qubits, higher baseline noise, and simpler circuit structures. Today’s 127-qubit superconducting machines are substantially different -- featuring better average fidelities, more complex qubit layouts, and greater heterogeneity across devices. As quantum applications scale and the number of required circuit executions increases, especially in variational and hybrid algorithms, the classical overhead of frequent transpilation becomes non-trivial~\cite{endo2021hybrid}. The cost of compiling thousands of circuit instances daily using Level 3 optimization -- each taking seconds of time to transpile, which is equivalent to or more than the actual execution time on the quantum computer~\cite{ravi2021quantum} -- can quickly dominate the overall workflow. 

These evolving realities motivate a careful re-examination of the assumptions underlying noise-adaptive transpilation. In this work, we study how transpilation strategies affect fidelity, performance variability, and classical overhead across a diverse set of algorithms executed on modern IBM quantum computers. Our goal is to understand not just how to optimize for short-term error reduction, but how to achieve performance stability and resource efficiency -- particularly as quantum algorithms scale beyond prototype problems.


\section{Experimental Methodology}
\label{sec:methodology}


\definecolor{headerblue}{HTML}{1F618D}     
\definecolor{stripegray}{HTML}{F2F2F2}    
\sisetup{table-number-alignment=center}
\newcolumntype{L}[1]{>{\raggedright\arraybackslash}p{#1}}  
\newcolumntype{C}[1]{>{\centering\arraybackslash}p{#1}}    
\begin{table}[t]
  \centering
  \caption{QASMBench~\cite{li2023qasmbench} algorithms used in our study.}
  \label{tab:algos}
  \rowcolors{2}{stripegray}{white}
  \begin{tabular}{
      L{0.8cm}                
      L{4.5cm}                
      C{0.8cm}                
      C{0.8cm}                
    }
    \toprule
    \rowcolor{pink!10}
    \color{black}\textbf{Algo.} 
      & \color{black}\textbf{Description} 
      & \color{black}\textbf{\#Qubits} 
      & \color{black}\textbf{\#Gates}\\
    \midrule
    \noalign{\renewcommand{\arraystretch}{1.0}}
    VQE   & Variational Quantum Eigensolver                  &  6 & 2282 \\
    HHL   & Harrow–Hassidim–Lloyd Algorithm                    &  7 &  689 \\
    QPE1  & Quantum Phase Estimation (Small)                   &  9 &  123 \\
    ADD1  & Quantum Adder (Small)                              & 10 &  142 \\
    SAT   & Satisfiability Circuit                             & 11 &  679 \\
    SECA  & Shor's Error Correction Algorithm                  & 11 &  216 \\
    GCM   & Generator Coordinate Method                        & 13 & 3148 \\
    MULT  & Quantum Multiplier                                 & 13 &   98 \\
    QPE2  & Quantum Phase Estimation (Large)                   & 15 &  311 \\
    DNN   & Quantum Deep Neural Network                        & 16 & 2016 \\
    QEC   & Quantum Error Correction Encoding                  & 17 &   53 \\
    ADD2  & Quantum Adder (Large)                              & 18 &  284 \\
    SQRT  & Square Root Circuit                                & 18 & 2300 \\
    QRAM  & Quantum Random Access Memory                       & 20 &   92 \\
    KNN   & Quantum $k$-Nearest Neighbors (Small)              & 25 &  230 \\
    SWAP  & SWAP Test Circuit                                  & 25 &  230 \\
    \bottomrule
  \end{tabular}
  \vspace{-1mm}
\end{table}

\noindent\textbf{Experimental Setup and Software:} We performed our experiments using IBM's quantum computing cloud services. Our implementation used IBM's open-source quantum computing framework Qiskit~\cite{aleksandrowiczqiskit} version 1.2.1, with Python 3.11.9 scripts for managing cloud job submissions and result retrieval. Bash shell scripts are used for automated job submissions, scheduled via crontab for timely execution. The experiments were conducted on six IBM quantum processors: \textit{ibm\_cusco}, \textit{ibm\_kyoto}, \textit{ibm\_nazca}, \textit{ibmq\_kyiv}, \textit{ibmq\_brisbane}, and \textit{ibmq\_sherbrooke}. Each quantum computer features an identical 127-qubit architecture and topography. All temporal data in our analysis references the UTC-5 (EST) time zone.

\vspace{2mm}

\noindent\textbf{Experiments Conducted:} We divide experiments into three groups, shown in Fig.\ref{fig:grouping}. The first group (Group-O3) reflects the conventional approach used in most quantum computing workflows: full noise-aware transpilation using optimization level 3 on the latest calibration data\cite{murali2019noise,tannu2019not,ash2020experimental}. This baseline is widely adopted in practice and assumed to yield the highest fidelity. The second group (Group-IM) evaluates whether randomizing the initial mapping (while retaining all routing optimizations) can yield comparable fidelity with reduced variability by avoiding repeated use of the same ``good'' qubits. This tests the hypothesis that uniform qubit usage may improve performance stability. The third group (Group-OL) investigates whether lower optimization levels (1 or 2) can match the fidelity of level 3 while significantly reducing classical transpilation time. Together, these three groups allow us to isolate the impact of mapping strategy and transpilation complexity on fidelity and performance variability. Each group was structured to decouple distinct factors: standard transpilation, daily noise adaptation, and stochastic remapping. This separation enables a controlled comparison of fidelity, variability, and runtime overhead across strategies.

Note: the three clusters in Group-O3 are all the same color in the figure as they run the exact same transpilation of the circuits thrice for statistical meaningfulness. The other groups can have randomization among the clusters due to random initial mapping and varying optimization levels. Hence, they have differently-color-coded clusters.  We detail how each group supports specific analysis in Sec.~\ref{sec:analysis}.

\vspace{2mm}

\noindent\textbf{Algorithms Evaluated:} We evaluate 16 algorithms listed in Table~\ref{tab:algos}. These algorithms are taken from the QASMBench~\cite{li2023qasmbench} benchmark suite. We selected these algorithms to represent a variety of diverse characteristics in terms of width and depth. Note: We cap the circuit width at 25 qubits because beyond that, the circuit produced uninterpretable results due to the noise on current quantum computers. Further, evaluating 6-25-qubit circuits on 127-qubit hardware allows us to investigate how transpilation strategies interact with hardware heterogeneity and qubit selection policies, especially in the context of large and sparsely connected topologies.

\vspace{2mm}

\noindent\textbf{Quantum Computers Used:} We ran our experiments on six 127-qubit IBM quantum computers: Brisbane, Cusco, Kyiv, Kyoto, Nazca, and Sherbrooke (limited use). All computers feature the Eagle r3 processor architecture. Experiments were run for all three groups on all computers, not necessarily at the same time, due to varying computer availabilities. At least 3-8 days of experiments were run on each computer. A total of 4958 circuits were run across all six computers over an experimentation period spanning approximately one month.

\vspace{2mm}

\noindent\textbf{Statistical Metrics used for Analysis:} To gauge the output fidelity of quantum runs, we must compare the ideal quantum simulation output and the noisy output when the circuits are run on real computers. Various metrics could be used to capture this output difference. Total Variation Distance (TVD), Hellinger Distance (HLD), Jensen-Shannon Distance (JSD), and Kullback-Leibler Divergence (KLD) are common measures~\cite{tannu2019ensemble,patel2022quest,das2019case}. These measures are all closely related; for instance, HLD bounds TVD and JSD bounds KLD~\cite{ding2023empirical,okamura2023metrization}. \textit{Thus, we select KLD for our study as it is frequently chosen for its interpretability in terms of information loss in the noisy output, making it a comprehensive metric for comparing quantum distributions.} For two probability distributions $P$ and $D$, their KLD is calculated as $D_{\text{KL}} = \sum_{i\in S} P(i) \log \frac{P(i)}{Q(i)}$, over output states $i\in S$. We also refer to KLD as ``output error'' and its inverse as ``output fidelity''~\cite{van2014renyi}.

We also use the Spearman correlation to measure the monotonic relationship between two variables using their rank orders, making it robust to outliers and non-linearities. It ranges from -1 (perfect negative correlation) to 1 (perfect positive correlation), with 0 indicating no monotonic relationship. It is given by $\rho = 1 - \frac{6 \sum d_i^2}{m(m^2 - 1)}$, where $d_i$ is the rank difference for each sample pair, and $m$ is the number of samples~\cite{wissler1905spearman}.


\section{Observations and Analysis}
\label{sec:analysis}

\begin{figure}[t]
    \centering
    \includegraphics[width=0.98\columnwidth]{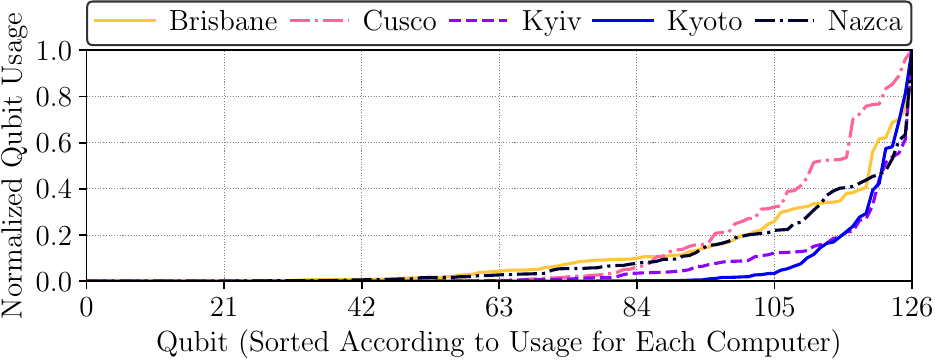}
    \vspace{1.5mm}
    \hrule
    \vspace{1.5mm}    \caption{Our analysis indicates that across all quantum computers, a small set of qubits gets used more frequently.}
    \vspace{-3mm}
    \label{fig:qubit_use}
\end{figure}

We organize our analysis into two categories: spatial effects, which capture how frequently and where circuits are mapped, and temporal effects, which consider when in time they are executed. Together, these analyses probe the core assumptions behind noise-adaptive transpilation, such as the need to always use ``good'' qubits and the necessity of continual re-transpilation over time.

\subsection{Spatial Observations and Analysis}

The first research question (RQ) we pose is about the frequency of qubit usage across the five quantum computers to examine how transpilers distribute computational load. This question is central to our broader investigation: if noise-aware transpilation always prioritizes qubits with the best noise properties~\cite {tannu2019not,patel2020experimental}, it may repeatedly map circuits to the same physical qubits. Such behavior not only reflects transpiler bias, but also has potential downstream effects -- especially as workloads scale up and those qubits become overutilized. 

\vspace{2mm}


\begin{table}[t]
    \centering
    \caption{Spearman correlation between qubit usage frequency and different qubit properties.}
    \label{tab:correls1}
    
    \definecolor{sigPos}{HTML}{9CC3E6}     
    \definecolor{weakPos}{HTML}{DEEBF7}    
    \definecolor{insig}{HTML}{F2F2F2}      
    \definecolor{weakNeg}{HTML}{FCDADA}    
    \definecolor{sigNeg}{HTML}{F8A9A9}     
    \definecolor{header}{HTML}{EFEFEF}     
    
    \newcommand{\ssig}{\textsuperscript{***}}  
    \newcommand{\msig}{\textsuperscript{**}}   
    \newcommand{\wsig}{\textsuperscript{*}}    
    \scalebox{0.88}{
        \begin{tabular}{l 
                        >{\raggedleft\arraybackslash}p{2.5cm} 
                        >{\raggedleft\arraybackslash}p{2.5cm}}
        \toprule
        \rowcolor{pink!10}
        \textbf{Property} & \textbf{Correlation} & \textbf{$p$-value} \\
        \midrule
        
        \rowcolor{cyan!10}
        Mean T1 Time & +0.08 & $6.73\times10^{-6}$ \\
        
        \rowcolor{cyan!10}
        Mean T2 Time & +0.18 & $1.25\times10^{-27}$ \\
        
        \rowcolor{white}
        Mean Qubit Frequency & -0.02 & $0.14$ \\
        
        \rowcolor{purple!10}
        Mean One-Qubit Gate Error & -0.19 & $1.27\times10^{-31}$ \\
        
        \rowcolor{purple!10}
        Mean Two-Qubit Gate Error & -0.28 & $4.10\times10^{-66}$ \\
        
        \rowcolor{purple!10}
        Mean Measurement Error & -0.18 & $4.12\times10^{-29}$ \\
        \bottomrule
        \end{tabular}
    }
\vspace{-1mm}
\end{table}


\begin{figure*}[t]
    \centering
    \includegraphics[width=0.19\textwidth]{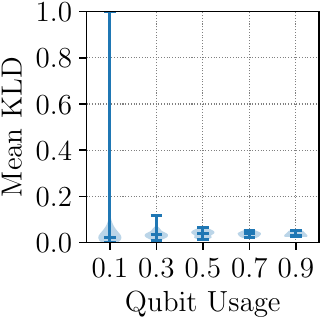}
    \hfill
    \includegraphics[width=0.19\textwidth]{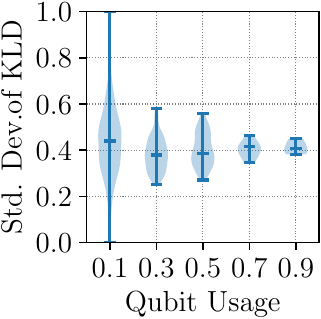}
    \hfill
    \includegraphics[width=0.19\textwidth]{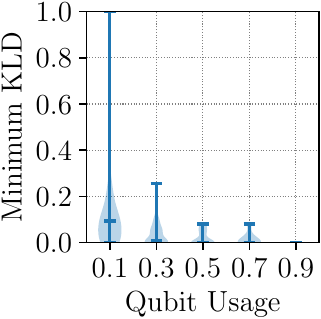}
    \hfill
    \includegraphics[width=0.19\textwidth]{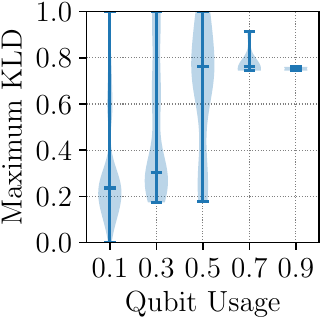}
    \hfill
    \includegraphics[width=0.19\textwidth]{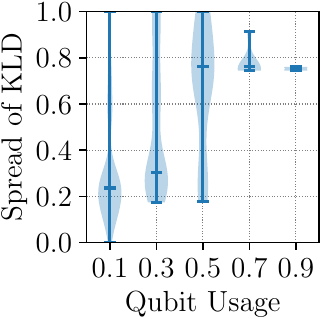}
    \vspace{1.5mm}
    \hrule
    \vspace{1.5mm}
    \caption{Relationship between qubit usage and different KLD metrics (exponentiated) demonstrates the strong correlation.}
    \vspace{-3mm}
    \label{fig:scatters}
\end{figure*}

\noindent\textbf{RQ \showRQcounter: Do some qubits get used more frequently than other qubits? What characteristics affect this frequency?}

\vspace{2mm}

Our analysis confirms that a small subset of qubits indeed gets used more frequently across all five quantum computers, as illustrated in Fig.~\ref{fig:qubit_use}. Following the Pareto Principle, about 80\% of the qubits get used less than 20\% of the time. This aligns with our hypothesis that the transpiler’s noise-aware optimizations tend to favor qubits with better noise characteristics, prioritizing those with lower error rates and more stable coherence times. Across the experiments, we observed that the same core subset of qubits repeatedly appeared in the transpiled circuits, regardless of the specific algorithms used.

To quantify this observation, we computed the correlation between qubit usage frequency and several qubit properties, including T1 and T2 times, gate error rates, and measurement error rates, as shown in Table~\ref{tab:correls1}. The analysis reveals a weak correlation with individual qubit properties -- for example, the negative correlation between the gate/measurement error rates and qubit usage is as expected because qubits with lower error rates are used more often. Similarly, qubits with higher coherence times are used more often; hence, that correlation is positive. The weakest correlation is with qubit frequency (a quantum property), which is also as expected, as frequency is not factored into making transpilation decisions. Note that the correlations with individual properties are weak because the optimization passes take into account multiple properties that could be conflicting (e.g., a qubit has high coherence times but also high gate errors). Nonetheless, the expected correlation signs confirm that the transpiler maps circuits to the highest-performing qubits available on each device.

\vspace{4mm}

\begin{insightbox}{}{}

Our results show that noise-aware transpilation heavily concentrates circuit mapping onto a limited set of qubits, resulting in a small fraction of qubits handling the majority of computational tasks. This constrained resource utilization reflects a load-imbalance bias toward minimal error rates, leaving most qubits unutilized. 

\end{insightbox} 

\vspace{2mm}

\begin{table}[t]
    \centering
    \caption{Spearman correlation between the qubit usage frequency and KLD properties of algorithms run on them.}
    \label{tab:correls2}
    
    \definecolor{sigPos}{HTML}{9CC3E6}     
    \definecolor{weakPos}{HTML}{DEEBF7}    
    \definecolor{insig}{HTML}{F2F2F2}      
    \definecolor{weakNeg}{HTML}{FCDADA}    
    \definecolor{sigNeg}{HTML}{F8A9A9}     
    \definecolor{header}{HTML}{EFEFEF}     
    
    \newcommand{\ssig}{\textsuperscript{***}}  
    \newcommand{\msig}{\textsuperscript{**}}   
    \newcommand{\wsig}{\textsuperscript{*}}    
    \scalebox{0.88}{
        \begin{tabular}{l 
                        >{\raggedleft\arraybackslash}p{2.5cm} 
                        >{\raggedleft\arraybackslash}p{2.5cm}}
        \toprule
        \rowcolor{pink!10}
        \textbf{Metric} & \textbf{Correlation} & \textbf{$p$-value} \\
        \midrule
        
        \rowcolor{cyan!10}
        Mean KLD & +0.39 & $6.28\times10^{-18}$ \\
        
        \rowcolor{cyan!10}
        Standard Deviation of KLD & +0.41 & $8.96\times10^{-20}$ \\
        
        \rowcolor{cyan!10}
        Maximum KLD & +0.72 & $7.59\times10^{-74}$ \\
        
        \rowcolor{purple!10}
        Minimum KLD & -0.71 & $7.78\times10^{-69}$ \\
        
        \rowcolor{cyan!10}
        Spread of KLD & +0.79 & $5.49\times10^{-97}$ \\
        \bottomrule
        \end{tabular}
    }
    \vspace{-1mm}
\end{table}

This result confirms that state-of-the-art transpilation policies prioritize fidelity by exploiting qubit heterogeneity. However, it also highlights an unintended consequence: qubits with slightly better metrics bear disproportionate workload, while most others remain underutilized. Whether this bias is justified -- or whether it limits robustness and scalability -- becomes a key question in our subsequent analysis.

\vspace{2mm}

\noindent\textbf{RQ \showRQcounter: Does the frequency of qubit usage affect the fidelity of the algorithms that are run on those qubits?}

\vspace{2mm}

We now investigate whether the concentrated use of a small subset of qubits, as shown in RQ1, leads to observable effects on algorithm fidelity. Our core hypothesis is that while noise-aware transpilation may improve short-term fidelity by selecting the ``best'' qubits, this strategy could result in volatility or instability as those qubits are overused, particularly in workloads requiring repeated executions. To evaluate this, we analyze the correlation between qubit usage frequency and several fidelity metrics derived from the KLD. Table~\ref{tab:correls2} reports the Spearman correlations between usage frequency and the mean, standard deviation, maximum, minimum, and spread of the KLD for circuits using those qubits.

Our findings, presented in Table~\ref{tab:correls2}, show a small positive correlation (0.39) between qubit usage frequency and the mean KLD, indicating that circuits mapped to frequently used qubits may experience higher average output divergence from ideal results. More significantly and surprisingly, the maximum KLD exhibits a strong positive correlation (0.72) with qubit usage, and the minimum KLD exhibits a strong negative correlation (-0.71) with qubit usage, suggesting that frequently used qubits are more prone to worst-case fidelity degradation. The correlation between qubit usage frequency and the KLD spread (maximum KLD - minimum KLD) is similarly high (0.79), which suggests that the fidelity variability, or the fluctuations around the mean, increase with higher usage.

Fig.~\ref{fig:scatters} further illustrates these relationships, with violin charts showing that as qubit usage increases, we observe higher KLD metrics, particularly in terms of maximum and spread. It is important to clarify what we are not claiming: this result does not prove that high-usage qubits undergo long-term physical degradation (e.g., from aging or permanent loss of coherence). In fact, our temporal analysis in later sections shows that compiled circuits can maintain fidelity across multiple calibration cycles. Instead, the takeaway is that aggressive reuse of select qubits leads to larger output variability in the short term. By output variability, we refer to the inconsistency in output distributions across repeated executions of the same circuit. Increased variability implies less predictable and less reliable quantum outputs. This is likely a result of dynamic cross-talk, thermal fluctuations, and other correlated noise sources that affect the ``best'' qubits more often simply because they are more used.

\vspace{4mm}

\begin{insightbox}{}{}

Our analysis shows that, as opposed to conventional belief~\cite{tannu2019ensemble,tannu2019not,murali2019noise,patel2020experimental}, noise-adaptive mapping that focuses on the ``best'' qubits may actually degrade fidelity. Frequently used qubits yield higher worst-case error and larger variability in fidelity. While noise-aware transpilation selects these qubits to minimize immediate error, our results suggest that this over-reliance may increase unpredictability and reduce robustness, especially for workloads sensitive to fidelity fluctuations.

\end{insightbox}

\vspace{2mm}


\begin{table}[t]
    \centering
    \caption{Spearman correlation between the qubit usage frequency and KLD properties of algorithms run on them when algorithms are mapped to random qubits.}
    \label{tab:correls3}
    
    \definecolor{sigPos}{HTML}{9CC3E6}     
    \definecolor{weakPos}{HTML}{DEEBF7}    
    \definecolor{insig}{HTML}{F2F2F2}      
    \definecolor{weakNeg}{HTML}{FCDADA}    
    \definecolor{sigNeg}{HTML}{F8A9A9}     
    \definecolor{header}{HTML}{EFEFEF}     
    
    \newcommand{\ssig}{\textsuperscript{***}}  
    \newcommand{\msig}{\textsuperscript{**}}   
    \newcommand{\wsig}{\textsuperscript{*}}    
    \scalebox{0.88}{
        \begin{tabular}{l 
                        >{\raggedleft\arraybackslash}p{2.5cm} 
                        >{\raggedleft\arraybackslash}p{2.5cm}}
        \toprule
        \rowcolor{pink!10}
        \textbf{Metric} & \textbf{Correlation} & \textbf{$p$-value} \\
        \midrule
        
        \rowcolor{cyan!10}
        Mean KLD & +0.27 & $6.41\times10^{-12}$ \\
        
        \rowcolor{purple!10}
        Standard Deviation of KLD & -0.27 & $2.54\times10^{-12}$ \\
        
        \rowcolor{cyan!10}
        Maximum KLD & +0.45 & $7.30\times10^{-33}$ \\
        
        \rowcolor{purple!10}
        Minimum KLD & -0.43 & $1.99\times10^{-30}$ \\
        
        \rowcolor{cyan!10}
        Spread of KLD & +0.47 & $4.28\times10^{-36}$ \\
        \bottomrule
        \end{tabular}
    }
    \vspace{-1mm}
\end{table}

\begin{figure}[t]
    \centering
    \includegraphics[width=0.98\columnwidth]{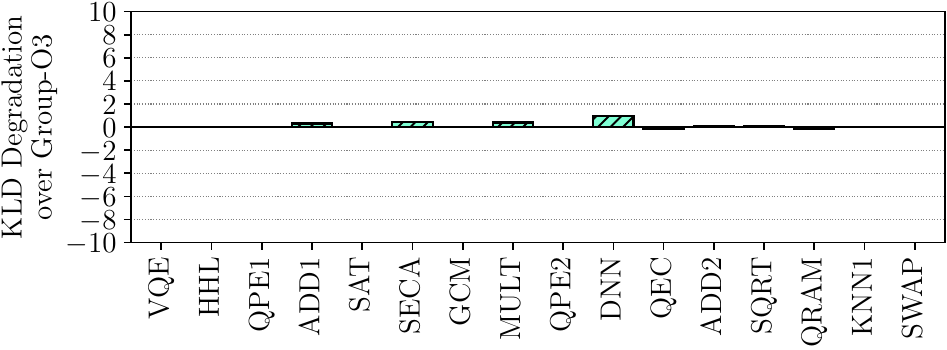}
    \vspace{1.5mm}
    \hrule
    \vspace{1.5mm}
    \caption{The average KLD difference for each algorithm when run with the optimal mapping vs. random mapping is low.}
    \vspace{-3mm}
    \label{fig:kld_diff}
\end{figure}

These findings make a compelling case for revisiting the goals of noise-aware transpilation. Rather than minimizing average error, an alternative objective might be to minimize variance or worst-case performance, especially for batch-executed or iterative quantum programs. This result thus sets the stage for RQ3, where we test whether evenly distributing circuit mappings across all qubits, regardless of their noise profiles, can mitigate variability while maintaining fidelity.

\vspace{2mm}

\noindent\textbf{RQ \showRQcounter: If the algorithms are distributed more evenly across different qubits, will we observe more stable results? Will the quality of the result decrease?}

\vspace{2mm}

To address RQ3, we tested whether distributing algorithms more evenly across different qubits (using uniform random mapping) rather than concentrating them on the ``best'' qubits would yield more stable fidelity without compromising accuracy. We analyze this using Group-IM, which applies the same transpilation optimizations as Group-O3 but introduces randomness in the initial qubit mapping. 

Table~\ref{tab:correls3} shows that the correlations between qubit usage frequency and KLD metrics drop consistently compared to Group-O3 (Table~\ref{tab:correls2}). Notably, the correlation with maximum KLD drops from 0.72 (Group-O3) to 0.45, and with KLD spread from 0.79 to 0.47. These substantial reductions indicate that a more balanced use of qubits significantly improves fidelity consistency. Even more importantly, the mean KLD remains largely unchanged (0.27 versus 0.39 in Group-O3), suggesting that the average fidelity does not suffer despite the lack of noise-aware mapping. This answers a key concern: while random mapping might appear naive, it still retains comparable average fidelity while offering improved robustness.

Why is the mean KLD similar despite omitting noise-aware optimization? A likely reason is that while noise-aware transpilation optimizes for the average case, it often crowds computation onto a small patch of the device. As a result, dynamic qubit wear may negate the theoretical gains from selecting low-error qubits. Random mapping, by contrast, spreads risk across the chip and may avoid correlated noise effects. Fig.~\ref{fig:kld_diff} confirms that the average fidelity of each algorithm under random mapping (Group-IM) is comparable to that achieved with noise-aware mapping (Group-O3) (KLD difference is $\le1$, which is considered small), suggesting that a more even workload distribution can maintain accuracy while reducing extreme fluctuations and worst-case errors. An additional advantage is that this also eliminates the classical computing overhead related to noise-adaptive mapping.

\vspace{4mm}

\begin{insightbox}{}{}

Our findings suggest that a simple random mapping strategy can maintain similar average fidelity to noise-adaptive transpilation while significantly reducing output variability and worst-case errors. This approach not only stabilizes performance but also removes the classical overhead associated with frequent noise-adaptive mapping for each device calibration (e.g., once daily). This insight also opens up the opportunity to co-locate multiple programs on the same computer (not currently practiced due to the limited number of ``good'' qubits), as they do not have to compete for the best qubits and can be assigned an arbitrary random section of the chip.

\end{insightbox}

\vspace{2mm}

These results also suggest new design principles for future transpilation strategies. Instead of always prioritizing lowest-error qubits, transpilers could include a diversity-aware term in their objective, trading small increases in expected KLD for reduced variance or more even qubit wear. For example, transpilers might randomize initial mappings within a subset of ``acceptable'' qubits or rotate mappings across calibrations to distribute usage over time. While our experiments use na\"ive random mapping, they motivate future work to formalize this trade-off between stability and short-term optimality.


\subsection{Temporal Observations and Analysis}

We now study the temporal effects of transpilation. The first natural question to ask is whether the classical overhead of transpilation is worth the effort to reduce the output error.

\vspace{2mm} 

\noindent\textbf{RQ \showRQcounter: Is the classical computation time spent on routing optimizations well worth the effort?}

\begin{figure}[t]
    \centering
    \includegraphics[width=0.98\columnwidth]{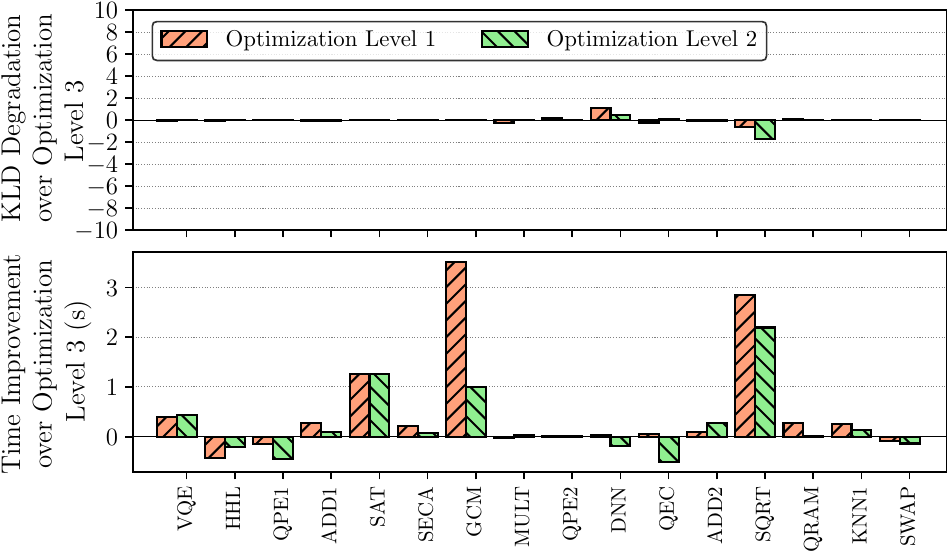}
    \vspace{1.5mm}
    \hrule
    \vspace{1.5mm}
    \caption{Optimization levels 1 and 2 achieve similar KLD values as optimization level 3, while achieving much lower transpilation time than optimization level 3.}
    \vspace{-3mm}
    \label{fig:rq4}
\end{figure}

\vspace{2mm} Fig.~\ref{fig:rq4} shows the impact of optimization levels on KLD and transpilation time. The upper plot displays the KLD difference between circuits compiled at optimization levels 1 and 2 versus level 3. The lower plot shows the corresponding time difference between level 3 and the lower levels. Our analysis reveals that the choice of optimization level has minimal impact on circuit performance. Circuits compiled with optimization levels 1 and 2 have comparable performance to those using optimization level 3, with the majority showing KLD degradation of zero and none exceeding a degradation of 2. Interestingly, in some cases, like the \textit{SQRT} result, lower optimization levels outperform optimization level 3. While the \textit{DNN} benchmark shows a slight advantage when using optimization level 3, this is an exception rather than the norm. These results suggest that the aggressive optimization strategies of level 3 are not always necessary to maintain or improve fidelity. We hypothesize that this trend is due to diminishing returns from fine-tuned qubit selection on newer devices, where hardware variability is lower and coherence times are more stable. Moreover, because optimization level 3 is sensitive to calibration noise at a single point in time, it may, in some cases, overfit to transient characteristics rather than improving long-term performance.

In parallel with our fidelity analysis, we also wanted to see how much transpilation time was sacrificed when using higher optimization levels. The lower plot of Fig.~\ref{fig:rq4} reveals that the majority of benchmarks (11 out of 16) demonstrate increased transpilation times when using optimization level 3 compared to levels 1 and 2. Furthermore, we observed that transpilation time overhead increases with algorithm size, making optimization level selection particularly crucial for larger circuits. Notably, certain algorithms, such as \textit{SAT}, \textit{GCM}, and \textit{SQRT}, exhibited significant temporal improvements, reducing transpilation time by up to three seconds per execution. While this seems modest in isolation, these delays quickly compound at scale. For instance, running 10,000 circuits in a variational algorithm could lead to hours of additional compile time. We further observed that transpilation time tends to scale superlinearly with algorithm size, suggesting that this cost will grow even more pronounced for future circuits approaching the full 127-qubit capacity of current devices.

\vspace{4mm}

\begin{insightbox}{}{}

Circuits compiled with lower optimization levels (1 or 2) achieve fidelity comparable to level 3, but with significantly lower transpilation time, especially as circuit size increases. Therefore, full noise-aware optimization should not be used indiscriminately; instead, lightweight strategies may be preferable for large or repeatedly executed workloads, such as variational quantum algorithms.

\end{insightbox}

\vspace{2mm}

The above insight propelled us to investigate the impact of the temporal characteristics of transpilation. Especially as IBM computers are calibrated once daily due to qubit drift~\cite{liu2023enabling} (around 12 am EST), we asked if it makes a difference if an algorithm is executed right after calibration or up to 24 hours after calibration. The implicit assumption is that the qubits may drift more significantly as more time passes after calibration, thus increasing the KLD. We test this assumption.

\vspace{2mm}

\begin{figure}[t]
    \centering
    \includegraphics[width=0.98\columnwidth]{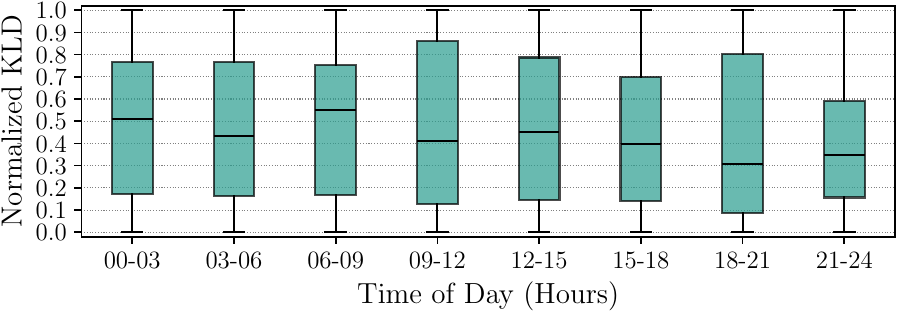}
    \vspace{1.5mm}
    \hrule
    \vspace{1.5mm}
    \caption{Temporal variability of KLD values across all algorithms shows no monotonic trend based on time of day.}
    \vspace{-3mm}
    \label{fig:rq5}
\end{figure}

\noindent\textbf{RQ \showRQcounter: Does the distance of the circuit execution time from the time of machine calibration impact the output fidelity?}

\vspace{2mm} To investigate how the time of execution during the day affects quantum circuit results, we employed Group-O3 (same circuit each time) to measure the relationship between time and KLD performance. Treating each (Algorithm, Computer) pair as an independent group, we normalized the KLD values within each group to a range of 0 to 1. The data was then grouped into three-hour bins and visualized using box plots, as shown in Fig.~\ref{fig:rq5}. Our analysis shows normalized KLD medians consistently ranging between 0.3 and 0.55 across all time periods. While we observed slightly higher variability during 18:00-21:00 and more stable performance during 21:00-24:00 (evidenced by a smaller interquartile range), these variations are minimal. The results suggest no significant correlation between execution time and output fidelity, indicating that quantum circuit performance remains stable through the day.

We also conducted a detailed correlation analysis for individual algorithms as shown in Table~\ref{tab:time_corr}. It presents the Spearman correlation coefficients between execution time and KLD values for each algorithm across all computers. The strongest correlations were found in VQE (0.36) and HHL (-0.36), both with p-values less than 0.001, yet these correlations are still relatively weak. Most algorithms showed even weaker correlations, with several (QRAM, SQRT, ADD2) showing correlations near zero. Interestingly, our analysis confirms that time-of-day effects remain insignificant not just on average but also at the individual algorithm level.

Importantly, this finding complements our earlier results: not only does random mapping yield more stable fidelity, but even noise-aware circuits compiled at the beginning of the day remain valid hours later. Together, these observations cast doubt on the assumption that circuit performance decays meaningfully with time since calibration. From a practical perspective, this opens the door to dramatically reducing transpilation frequency without sacrificing performance.


\begin{table}[t]
    \centering
    \caption{Spearman correlation between time of day and KLD values for each algorithm shows weak correlation.}
    \label{tab:time_corr}
    
    \definecolor{sigPos}{HTML}{9CC3E6}     
    \definecolor{weakPos}{HTML}{DEEBF7}    
    \definecolor{insig}{HTML}{F2F2F2}      
    \definecolor{weakNeg}{HTML}{FCDADA}    
    \definecolor{sigNeg}{HTML}{F8A9A9}     
    \definecolor{header}{HTML}{EFEFEF}     
    
    \newcommand{\ssig}{\textsuperscript{***}}  
    \newcommand{\msig}{\textsuperscript{**}}   
    \newcommand{\wsig}{\textsuperscript{*}}    
    \scalebox{0.9}{
        \begin{tabular}{>{\columncolor{purple!10}}l
                        >{\columncolor{pink!10}\raggedleft\arraybackslash}p{0.6cm} 
                        >{\columncolor{cyan!10}\raggedleft\arraybackslash}p{1.7cm}
                        >{\columncolor{purple!10}}l
                        >{\columncolor{pink!10}\raggedleft\arraybackslash}p{0.6cm} 
                        >{\columncolor{cyan!10}\raggedleft\arraybackslash}p{1.7cm}}
        \toprule
        \rowcolor{pink!10}
        \textbf{Algorithm} & \textbf{Corr.} & \textbf{$p$-value} & \textbf{Algorithm} & \textbf{Corr.} & \textbf{$p$-value} \\
        \midrule
        
        VQE & +0.36 & $8.78\times10^{-4}$ & 
        QPE2 & +0.11 & $3.21\times10^{-1}$ \\
        
        HHL & -0.36 & $9.04\times10^{-4}$ & 
        DNN & -0.09 & $4.24\times10^{-1}$ \\
        
        QPE1 & +0.34 & $1.84\times10^{-3}$ & 
        QEC & -0.08 & $4.78\times10^{-1}$ \\
        
        ADD1 & -0.34 & $2.13\times10^{-3}$ & 
        ADD2 & +0.05 & $6.36\times10^{-1}$ \\
        
        SAT & +0.29 & $9.25\times10^{-3}$ & 
        SQRT & -0.03 & $7.71\times10^{-1}$ \\
        
        SECA & -0.28 & $1.22\times10^{-2}$ & 
        QRAM & +0.01 & $9.57\times10^{-1}$ \\
        
        GCM & -0.25 & $2.33\times10^{-2}$ & 
        KNN & +0.14 & $2.11\times10^{-1}$ \\
        
        MULT & -0.20 & $6.87\times10^{-2}$ & 
        SWAP & -0.14 & $2.13\times10^{-1}$ \\
        \bottomrule
        \end{tabular}}
    \vspace{-1mm}
\end{table}

\vspace{4mm}

\begin{insightbox}{}{}

Our empirical findings indicate that the time elapsed since calibration -- whether minutes or up to 24 hours -- has minimal impact on output fidelity, i.e., there is no statistically significant relationship. This suggests that concerns about qubit drift post-calibration~\cite{liu2023enabling} may be overstated. This finding supports a shift toward fewer daily transpilation passes in production quantum workflows.

\end{insightbox}

\vspace{2mm}

Building on our hour-of-the-day analysis, we now ask whether transpiled circuits can be used across multiple days, even as the calibration data changes. This question probes the heart of current transpilation workflows, which assume that circuits must be recompiled every day using the most recent calibration. If this assumption does not hold, the implications for system efficiency and large-scale workloads are substantial.

\vspace{2mm}

\noindent\textbf{RQ \showRQcounter: Does a circuit mapped, routed, and optimized for one calibration cycle perform just as effectively for other calibration cycles without needing to be transpiled again?}

\vspace{2mm}

\begin{figure}[t]
    \centering
    \includegraphics[width=0.98\columnwidth]{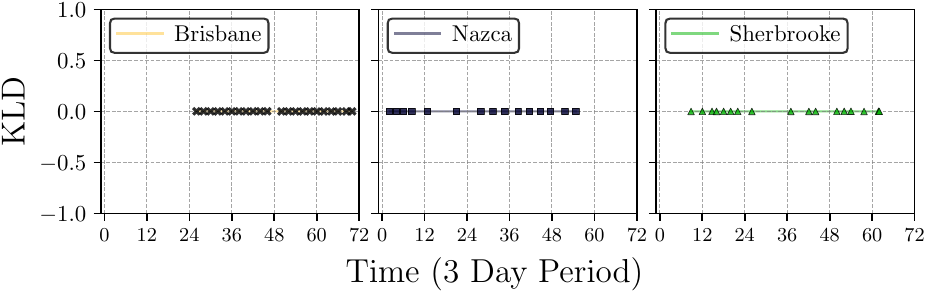}
    \vspace{1.5mm}
    \hrule
    \vspace{1.5mm}
    \caption{Results for the GCM algorithm on three computers over a duration of 72 hours show little change in KLD.}
    \vspace{-3mm}
    \label{fig:rq6}
\end{figure}

We examined the KLD values of the same circuit across three-day periods on three computers for a representative benchmark shown in Fig.~\ref{fig:rq6}. Due to varying queue times when accessing these quantum computers, the data points shown in the graphs have different times during which they end up being executed. The results show remarkably consistent KLD values near 0 across 72 hours for each computer. As shown in Fig.~\ref{fig:rq6}, the output fidelity, measured via KLD, remained remarkably stable across the 72-hour period. While minor fluctuations are expected due to queueing noise and machine load, there was no clear upward drift or consistent degradation trend. Practically, this opens the door to transpilation reuse — a simple optimization where a single transpilation output is cached and reused over time. This strategy could drastically reduce classical preprocessing cost, especially for variational algorithms that execute the same circuit thousands of times.

\vspace{4mm}

\begin{insightbox}{}{}

Circuits compiled once using calibration data from a single day retain stable fidelity across multiple calibration cycles, showing no observable performance degradation. This challenges the conventional wisdom that daily re-transpilation is necessary, and suggests that transpilation reuse may offer a practical path toward reducing compilation overhead for iterative quantum workloads~\cite{preskill2022quantum}.

\end{insightbox}

\vspace{2mm}

Overall, our study redefines best practices in noise-adaptive transpilation, showing that frequent re-transpilation may not be essential to maintain fidelity on the latest superconducting quantum devices. This work advances efficient, scalable quantum workflows, especially for resource-intensive algorithms.
\section{Related Work}
\label{sec:related_work}

Transpilation for noise-aware optimization has been a standard approach for improving quantum circuit performance on noisy quantum computing devices. As quantum circuits must be adapted to the error rates and hardware requirements of specific technologies and noise behaviors of different quantum computers, a large body of prior art has focused on developing transpilation strategies to improve fidelity, reduce error accumulation, and adapt to changing hardware conditions~\cite{endo2021hybrid,tannu2019ensemble,tannu2019not,murali2020software,zulehner2019compiling,liu2023tackling,tan2023compiling,patel2020ureqa,patel2020veritas}.

Early efforts in this space focused on noise-adaptive circuit mapping. Murali et al.~\cite{murali2019noise} proposed a compilation approach that adapts qubit placement to minimize the impact of noise by routing operations through more reliable qubits. Similarly, Tannu and Qureshi~\cite{tannu2019mitigating} developed dynamic strategies to mitigate qubit-specific error rates, showing that careful qubit allocation and scheduling can significantly improve output fidelity.  Li et al.\cite{li2019tackling} explored compilation techniques that strike a better balance between fidelity improvement and transpilation time, an issue that becomes critical as algorithm size increases. More recent work by Das et al.\cite{das2021adapt} extended this direction by proposing adaptive transpilation strategies that respond to evolving noise profiles.

Beyond optimization methods, several works have sought to empirically characterize the behavior of quantum hardware and understand how noise impacts transpilation effectiveness~\cite{ravi2021quantum,liu2020reliability,patel2020experimental,ash2020experimental,dahlhauser2021modeling}. For instance, Ash-Saki et al.~\cite{ash2020experimental} show that mitigating crosstalk through noise-aware transpilation can enhance fidelity, but it requires updated calibration data. Ravi et al.~\cite{ravi2021quantum} emphasize the need for continual noise-aware transpilation to counteract dynamic noise changes, especially in cloud-based systems with high user contention. On the other hand, Dahlhauser et al.~\cite{dahlhauser2021modeling} model noisy circuits by decomposing them into smaller subcircuits for dynamic tuning, which reflects a granular approach to transpilation.

\textit{While these works underscore the importance of tailoring quantum circuits to hardware noise, they were primarily conducted during an earlier phase of quantum computing, typically using devices with fewer qubits and higher baseline noise levels.} As quantum processors have grown and improved, the question arises whether these strategies continue to provide significant benefits. \textit{Our work directly addresses this question through a large-scale empirical study on 127-qubit IBM quantum computers, re-evaluating the effectiveness and cost of noise-aware transpilation in light of modern hardware.}


\section{Conclusion}
\label{sec:conclusion}

This work revisited a long-standing assumption in quantum computing: that noise-aware transpilation must be performed frequently to ensure high-fidelity execution. Through a large-scale empirical study, we found that this assumption does not always hold. Our results showed that noise-adaptive transpilation often concentrates execution on a small set of qubits, leading to instability and increased output variability. We demonstrated that random mapping strategies, though seemingly naive, can provide better load balancing and comparable fidelity while incurring lower compilation overhead.

Moreover, we found that circuit performance remains consistent not only throughout a 24-hour calibration cycle but also across multiple calibration cycles, enabling the reliable reuse of a single compiled circuit. These insights offer a new perspective on transpilation: it is not just about optimizing for transient noise characteristics, but about sustaining stable performance while minimizing classical costs. As quantum workloads scale up -- particularly variational quantum algorithms -- the ability to avoid frequent re-transpilation could dramatically reduce the classical overhead. We recommend a shift in best practices toward transpilation strategies that are both noise-aware and resource-efficient, especially as we approach the early fault-tolerant era of quantum computing.

\section*{Acknowledgement}

We would like to thank the anonymous ICCAD reviewers, whose feedback helped improve this work. This work was supported by Rice University, the Rice University George R. Brown School of Engineering and Computing, and the Rice University Department of Computer Science. This work was supported by the DOE Quantum Testbed Finder Award DE-SC0024301, the Ken Kennedy Institute, and Rice Quantum Initiative, which is part of the Smalley-Curl Institute.

This work was also supported by the following NSF grants: CSR-2402328, CAREER-2338457, CSR-2406069, CSR-2323100, and HRD-2225201, as well as by the DOE Office of Science User Facility, supported by the Office of Science of the U.S. Department of Energy under Contract No. DE-AC02-05CH11231, using NERSC award DDR-ERCAP0034716. We acknowledge the use of IBM Quantum services for this work. The views expressed are those of the authors, and do not reflect the official policy or position of IBM or the IBM Quantum team.

\balance

\bibliographystyle{plain}
\bibliography{main}

\end{document}